\documentclass{article} 
\usepackage{graphicx}
\graphicspath{{converted_graphics/}}
\begin{document}

\begin{center}
{\LARGE Measurements of Snow Crystal Growth Dynamics}\vskip6pt

{\LARGE in a Free-fall Convection Chamber}\vskip16pt

{\Large Kenneth G. Libbrecht}\footnote{%
kgl@caltech.edu. For the latest version of this and related papers, see

http://www.its.caltech.edu/\symbol{126}atomic/publist/kglpub.htm}{\Large ,
Helen C. Morrison, and Benjamin Faber}\vskip4pt

{\large Department of Physics, California Institute of Technology}\vskip-1pt

{\large Pasadena, California 91125}

\vskip18pt

\hrule \vskip1pt \hrule
\vskip 14pt
\end{center}

\noindent \textbf{Abstract. We present a series of experiments investigating
the growth of ice crystals from water vapor in the presence of a background
gas. We measured growth dynamics at temperatures ranging from -2 C to -25 C,
at supersaturations between 0.5 and 30 percent, and with background gases of
nitrogen, argon, and air at a pressure of one bar. We compared our data with
numerical models of diffusion-limited growth based on cellular automata to
extract surface growth parameters at different temperatures and
supersaturations. These data represent a first step toward obtaining
precision ice growth measurements as a function of temperature,
supersaturation, background gas pressure and gas constituents. From these
investigations we hope to better understand the surface molecular dynamics
that determine crystal growth rates and growth morphologies.}

\section{\noindent Introduction}

The formation of complex structures during solidification often results from
a subtle interplay of nonequilibrium, nonlinear processes, for which
seemingly small changes in molecular dynamics at the nanoscale can produce
large morphological changes at all scales. One popular example of this
phenomenon is the formation of snow crystals, which are ice crystals that
grow from water vapor in an inert background gas. Although this is a
relatively simple physical system, snow crystals display a remarkable
variety of columnar and plate-like forms, and much of the phenomenology of
their growth remains poorly understood \cite{libbrechtreview}.

Recent experimental and theoretical work suggests that surface impurities
play an essential role in determining snow crystal growth rates and
morphologies under normal atmospheric conditions \cite{impurities}. To
investigate this further we need precision measurements of snow crystal
growth dynamics over a range of conditions, especially as a function of
impurity type and concentration within a background gas. We have constructed
a free-fall convection chamber for making such measurements \cite{chamber},
and we describe here our first data using this apparatus. The data were
mainly taken in ordinary laboratory air at a pressure of one bar and over a
range of temperatures and supersaturations. We began our experiments with
these conditions in order to establish a baseline of growth measurements
which can be used to determine optimal experimental strategies for
investigating ice growth in the presence of different impurities.

There have been a number of previous measurements of ice crystal growth
under similar conditions \cite{yamashita, kobayashi, ryan, takahasi, libbyu}%
, but the measurements presented here are a substantial improvement in terms
of scope and overall precision. To date there have been essentially no
systematic investigations of the effects of surface impurities on ice growth
dynamics \cite{libbrechtreview}.

\begin{figure}[tbp] % float placement: (h)ere, page (t)op, page (b)ottom, other (p)age
  \centering
  % file name: C:/Documents and Settings/Kenneth Libbrecht/My Documents/aatempfold/HelenObs/overviewx.jpg
  \includegraphics[width=4.5in,keepaspectratio]{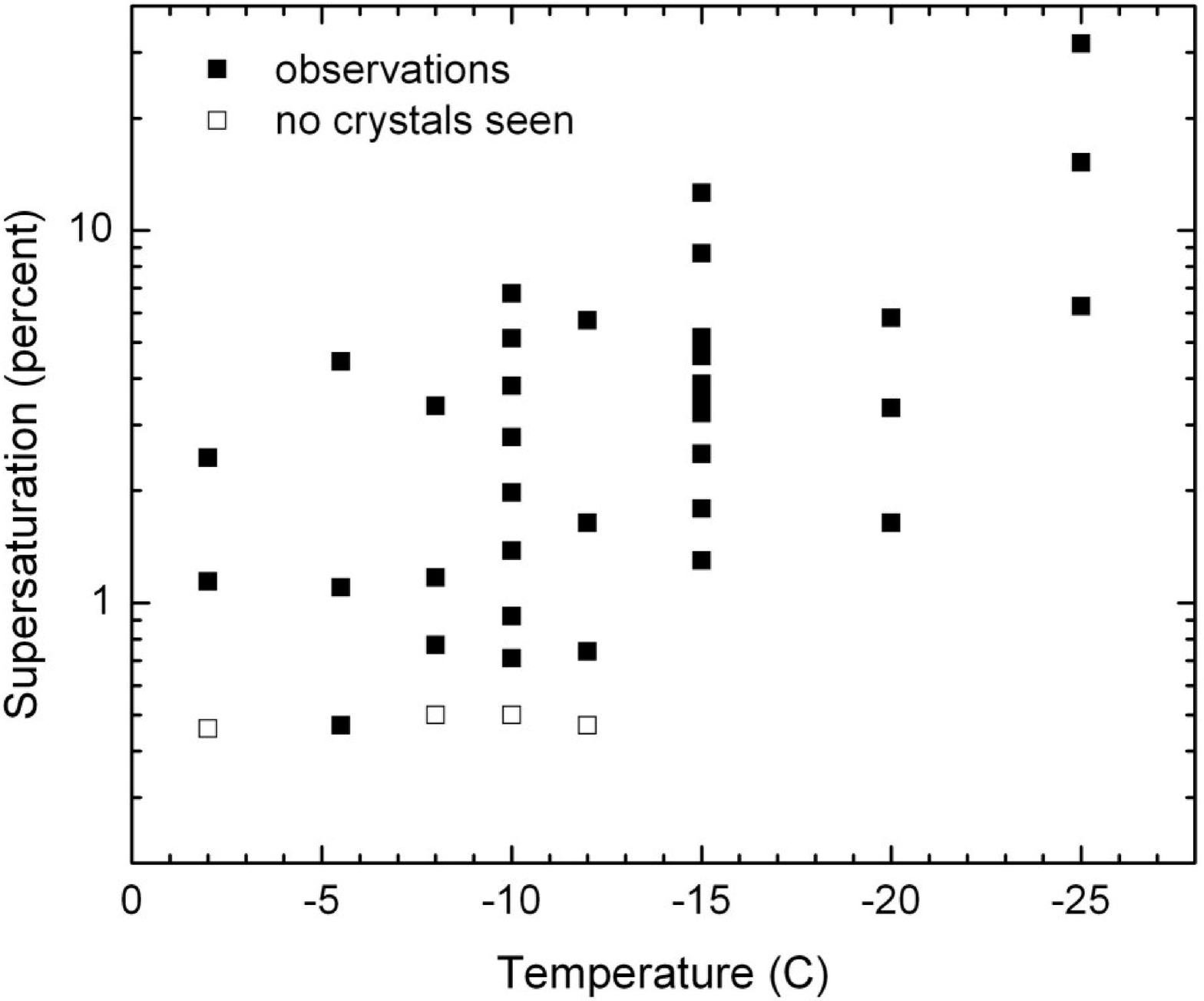}
  \caption{Graphical summary of the
different temperatures and supersaturations at which we made ice crystal
growth measurements for the results presented here. Open symbols show
conditions at which we were unable to grow crystals.}
  \label{overview}
\end{figure}

\section{Observations and Modeling}

Our experiments were performed in an ice crystal growth chamber that we
described in \cite{chamber}. The chamber was chilled to a desired
temperature and filled with a background gas at a pressure of one bar. We
used a heated reservoir filled with deionized water inside the chamber to
produce a known water vapor supersaturation via evaporation and convective
mixing. A number of crystals were nucleated and allowed to grow for several
minutes while in free-fall inside the chamber. As these crystals fell (or
were carried by convective currents) to a substrate at the bottom of the
chamber, we measured their size and thickness using a combination of optical
imaging and broad-band interferometry \cite{chamber}. From observations of a
large number of crystals we obtained the average crystal dimensions as a
function of growth time under conditions of known temperature and
supersaturation, as well as some sense of the distribution of these
quantities. Because of outgassing from the chamber walls and other sources,
we expect that the background gas included a number of unknown impurities at
the part-per-million level. We now believe that these impurities have a
substantial effect on the ice growth dynamics \cite{impurities}, which we
will be investigating in future experiments.

\begin{figure}[tbp] % float placement: (h)ere, page (t)op, page (b)ottom, other (p)age
  \centering
  % file name: C:/Documents and Settings/Kenneth Libbrecht/My Documents/aatempfold/HelenObs/d2x.jpg
  \includegraphics[width=4.5in,keepaspectratio]{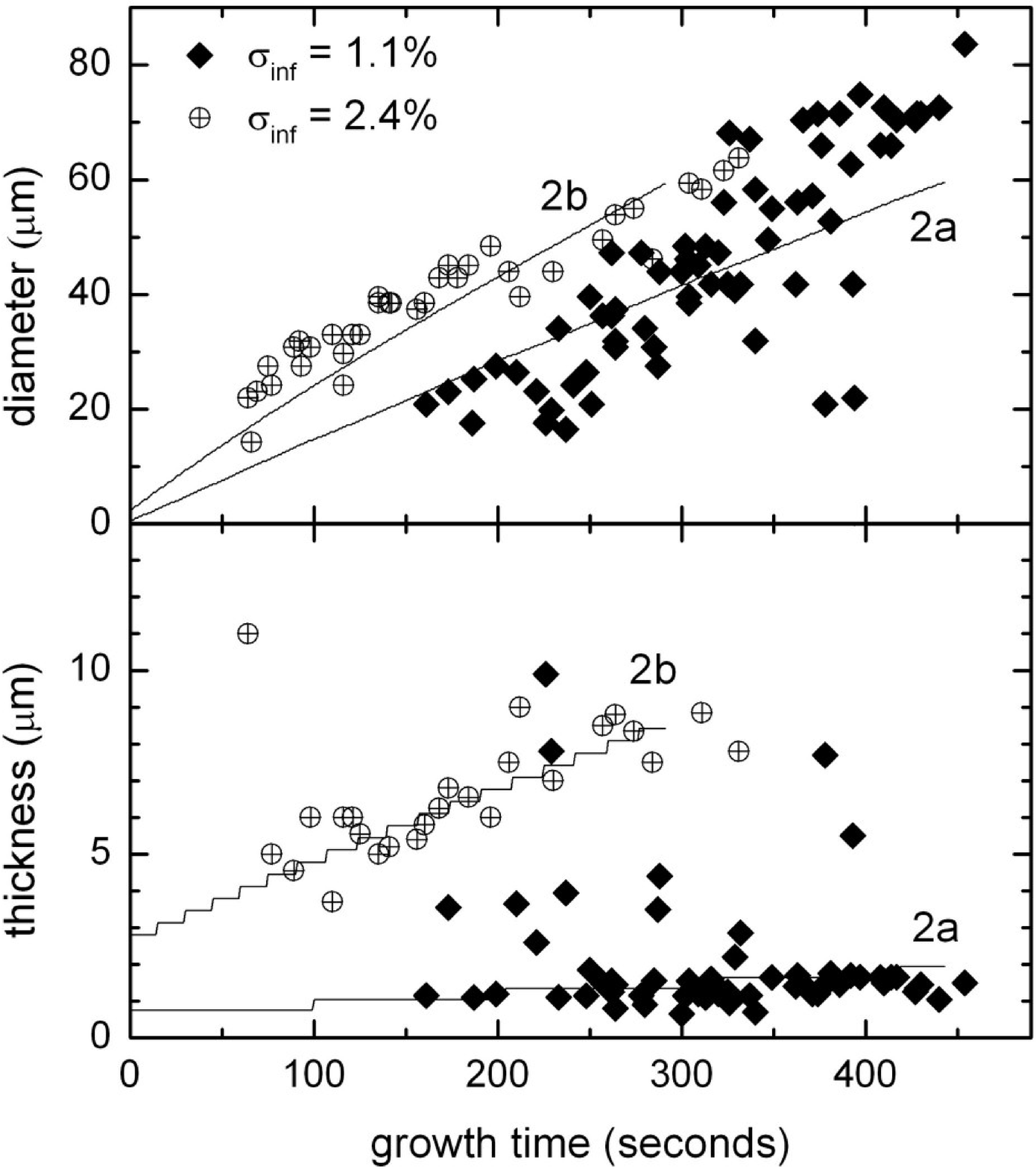}
  \caption{Ice growth data taken at -2 C at supersaturations of 1.1\% and
2.4\%. Each point represents the measurement of a single plate-like crystal.
The labeled lines show model calculations using the parameters given in the
Appendix.}
  \label{d2}
\end{figure}

Figure \ref{overview} shows the different temperatures and supersaturations
at which we made observations in ordinary laboratory air at a pressure of
one bar. Before each run we cleaned the chamber by heating it to
approximately 40 C and slowly replacing the air inside the chamber with new
filtered air over a period of typically 20-40 hours. This gentle bake
reduced solvent vapors and other impurities in the air to approximately the
same density as in the outside laboratory air. For growth in other gases,
the chamber was pumped out and new gas was introduced prior to data taking.

For measurements at temperatures above -15 C, we first cooled the growth
chamber to -15 C for several hours before adding room-temperature water to
the reservoir. Evaporation from the added water quickly coated the chamber
walls with a thin layer of frost (and not supercooled droplets) to provide
well-known boundary conditions for producing a known supersaturation inside
the chamber. Following this we continuously nucleated a large number of
crystals for several minutes before warming the chamber to the desired set
temperature for data taking.

Most of the crystals we measured were small ($< 100$ $\mu $m in
size) with relatively simple morphologies, and many were simple hexagonal
prisms. Depending on temperature, the prism morphologies ranged from slender
columns to thin plate-like crystals. Where possible we measured the plate
thickness (or column length) as well as an overall crystal diameter. The
latter was defined as the distance between opposing prism facets of a
hexagonal prism. For the case of triangular crystals or other non-hexagonal
morphologies, we used a rough estimate of the equivalent hexagonal diameter.

We write the growth velocity normal to the surface in terms of the
Hertz-Knudsen formula \cite{libbrechtreview} 
\begin{equation}
v_{n}=\alpha v_{kin}\sigma _{surf}  \nonumber
\end{equation}%
where $v_{kin}$ is a temperature-dependent kinetic velocity derived from
statistical mechanics, $\sigma _{surf}\ $is the water vapor supersaturation
just above the growing surface, and $\alpha $ is the condensation
coefficient, which contains the surface physics that governs how water
molecules are incorporated into the ice lattice, collectively known as the
attachment kinetics. If molecules striking the surface are immediately
incorporated into it, then $\alpha =1;$ otherwise $\alpha \leq 1$. For ice
crystals growing in an inert background gas, diffusion of water molecules to
the ice surface often significantly impedes the growth \cite{libbrechtreview}%
, so that $\sigma _{surf}<\sigma _{\infty }.$

We used cellular automata to numerically model our data \cite%
{libbrechtmodel, gg} in order to extract $\alpha $ values from the
measurements. We used a 2D approximation in which a hexagonal prism is
approximated by a simple right cylinder, so the six prism facets are
replaced by a single cylindrical \textquotedblleft facet\textquotedblright\ 
\cite{libbrechtmodel}. We assumed constant values of $\alpha _{basal}$ and $%
\alpha _{prism}$ (i.e., with no dependence on $\sigma _{surf})$ that were
adjusted so that the crystal dimensions as a function of time matched our
measurements. The initial crystals in our models were typically small
spheres of input diameter $D_{0}.$ The initial shape was usually
unimportant, however, since the model crystals always grew out into faceted
shapes in a short time. For the small crystals measured, the morphologies
were relatively simple and $\sigma _{surf}$ was found to be fairly close to $%
\sigma _{\infty }.$ Thus we expect that modeling uncertainties in our
extracted $\alpha _{basal}$ and $\alpha _{prism}$ values were relatively
small. We believe that the most significant uncertainty in our current
investigations stems from the unknown levels of impurities in the background
gas, a topic we intend to investigate at length in future experiments.

Surface tension effects may have played some role in our measurements, but
only at the lowest supersaturations. The equilibrium vapor pressure above a
static spherical crystal is approximately%
\[
c_{eq}\approx c_{sat}\left( 1+\frac{2\delta }{R}\right) 
\]%
where $c_{sat}$ is the equilibrium vapor pressure above a flat surface, $%
\delta \approx 1$ nm is a surface tension term \cite{libbrechtreview} and $R$
is the crystal radius. From this we see that the supersaturation $\sigma
=(c-c_{sat})/c_{sat}$ must be greater than $\sigma _{\min }\approx 2\delta
/R $ for growth to occur. Put another way, the crystal size must be greater
than 2$R_{\min }\approx 4\delta /\sigma _{surf}$ or a crystal will not grow.
In some cases presented below we observed $\sigma _{surf}$ as low as 0.5
percent, implying $2R_{\min }\approx 0.8$ $\mu $m. Although we sometimes
observed crystals of approximately this thickness at low supersaturations,
in most cases the surface tension term was negligible. Thus we did not
include surface tension effects in our numerical models.

\section{Results}

Unless noted otherwise, the data presented in this section are from crystals
grown in a background of ordinary laboratory air at a pressure of one bar.
For some runs we evacuated the chamber several times while cold and filled
it to one bar with bottled nitrogen or argon gas, but noticed no change in
the crystal morphologies or growth rates. The temperature and
supersaturation inside the growth chamber were determined as described in 
\cite{chamber}. We used either nitrogen or argon gas for nucleation, as we
observed that both gases yielded indistinguishable results.

\subsection{T = -2 C}

Figure \ref{d2} shows data taken at -2 C. About 90\% of the crystals
observed were plate-like, with the remainder being columnar or irregularly
shaped crystals. In the figure we have included only well-formed plate-like
crystals. Triangular plates (see \cite{fieldguide} for example pictures)
were especially common at -2 C, accounting for approximately 15\% of
measured crystals at $\sigma _{\infty }=1.1\%$ and 7\% at $\sigma _{\infty
}=2.4\%.$ Lines in the figure show model calculations with input parameters
given in the Appendix.

At 1.1\%, note that the crystal diameters increase with time more rapidly
than the model would suggest at the later times, and that the thicknesses do
not increase substantially with time. We believe some of this behavior may
be due to a systematic sampling effect. We observed crystals as they fell
onto our substrate, and crystals with the highest air resistance would tend
to remain aloft for the longest time inside the chamber. This factor may
account for the preponderance of especially large, thin plates seen at later
times in Figure \ref{d2}. Note this systematic sampling error affects our
data most strongly at later times after most crystals have already fallen.

\subsection{T = -5.5 C}

\begin{figure}[tbp] % float placement: (h)ere, page (t)op, page (b)ottom, other (p)age
  \centering
  % file name: C:/Documents and Settings/Kenneth Libbrecht/My Documents/aatempfold/HelenObs/d5nucleationx.jpg
  \includegraphics[width=4.5in,keepaspectratio]{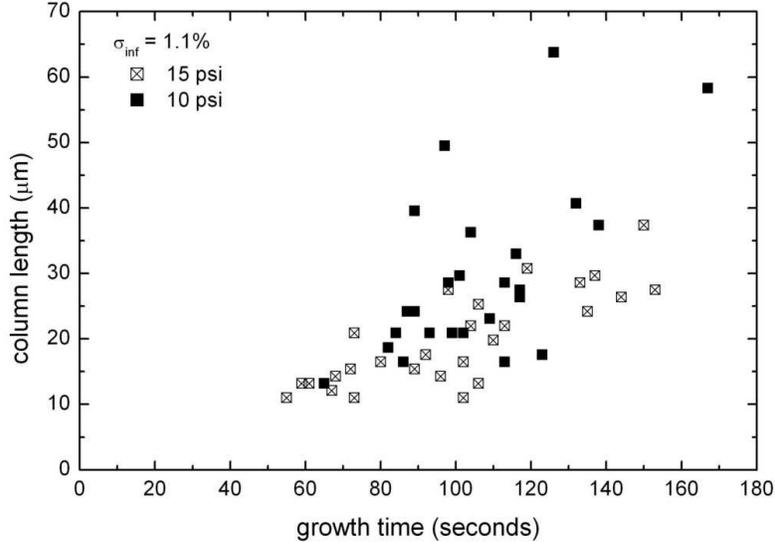}
  \caption{A comparison of crystals
produced at -5.5 C using two different nucleation pressures (see 
\protect\cite{chamber}). The higher pressure yielded more crystals, which
then competed for the available water vapor in the chamber and did not grow
as rapidly.}
  \label{d5nucleation}
\end{figure}

Figure \ref{d5nucleation} shows a test we performed at -5.5 C to look at
systematic effects from competition between growing crystals. If a large
number of crystals are nucleated inside the chamber at one time, their
growth lowers the supersaturation and reduces the final sizes of the
crystals. As a practice we decreased the nucleation pressure to within 1-2
psi of the minimum pressure required to produce crystals. We believe this
resulted in a sufficiently small number of crystals that systematic effects
in the measured crystal sizes from water vapor competition were less than
20\%.

\begin{figure}[tbp] % float placement: (h)ere, page (t)op, page (b)ottom, other (p)age
  \centering
  % file name: C:/Documents and Settings/Kenneth Libbrecht/My Documents/aatempfold/HelenObs/d5x.jpg
  \includegraphics[width=4.5in,keepaspectratio]{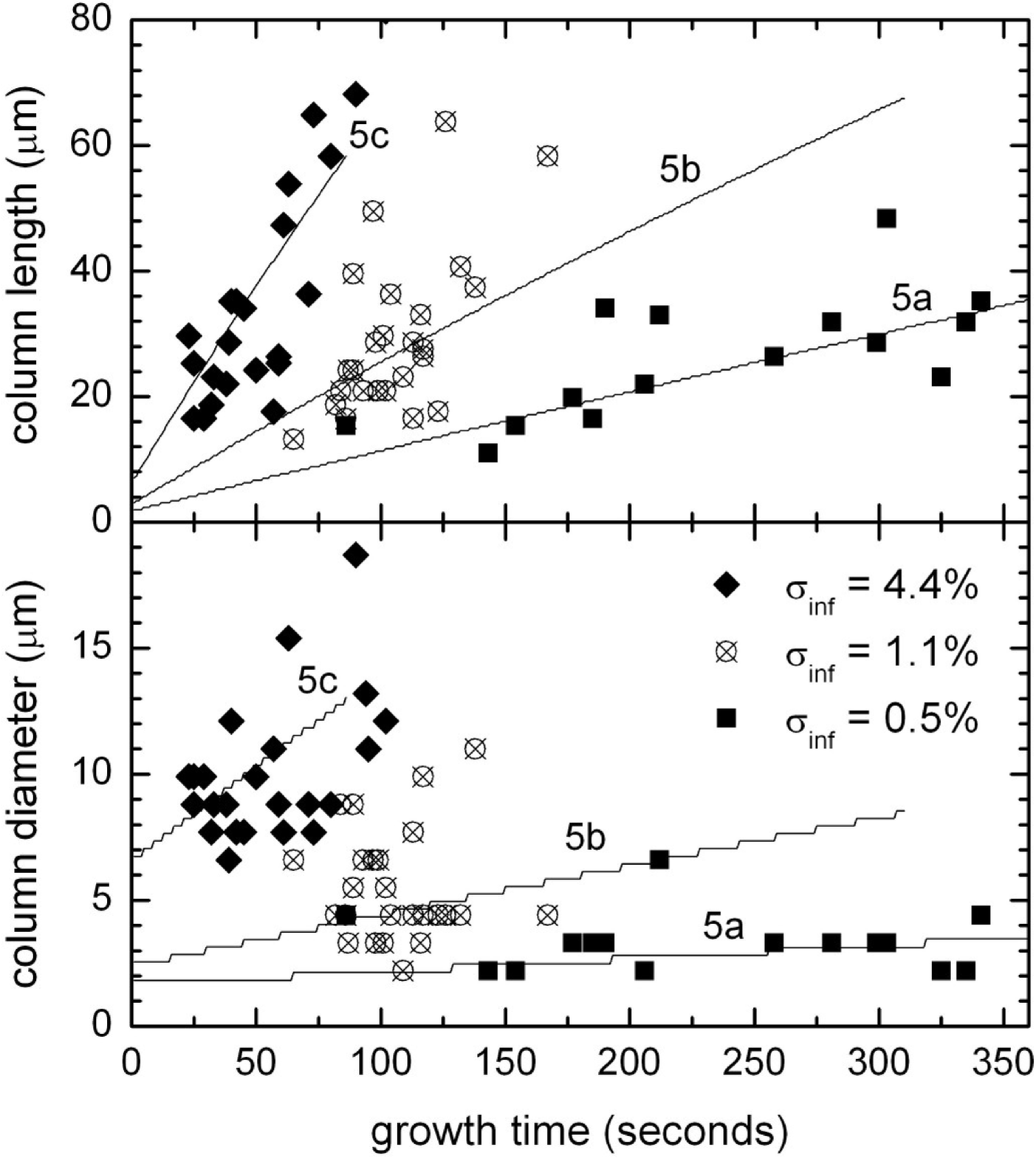}
  \caption{Ice growth data taken
at -5.5 C and several different supersaturations, along with model
calculations using the parameters in the Appendix.}
  \label{d5}
\end{figure}

Figure \ref{d5} shows data taken at -5.5 C along with model calculations.
Crystals observed at $\sigma _{\infty }=4.4\%$ were typically more
structured than at lower supersaturations, with hollow basal surfaces at
later times. Our measurements refer to overall crystal sizes, not including
hollows. Generally less structure was seen in our numerically modeled
crystals. The current data are in reasonable agreement with Libbrecht and Yu 
\cite{libbyu}, although in Figure \ref{d5} we show all crystals observed,
not just the largest ones, so the sizes here are somewhat smaller than in 
\cite{libbyu}.

In Figure \ref{d5} we see that the initial column diameter (at $t=0$ in the
model calculations) appears to increase with increasing supersaturation. The
origin of this is unknown at present, but we see similar trends in much of
the data presented here, particularly at temperatures between -5.5 C and -10
C. It seems unlikely that nucleation dynamics are responsible for this
effect, and we suspect that the crystal growth is especially fast at early
times, indicating that $\alpha $ is initially large and then is reduced to
some nearly constant value after a short period of time. The buildup of
surface impurities on an initially clean crystal may be responsible for this
behavior.

\begin{figure}[tbp] % float placement: (h)ere, page (t)op, page (b)ottom, other (p)age
  \centering
  % file name: C:/Documents and Settings/Kenneth Libbrecht/My Documents/aatempfold/HelenObs/d8histx.jpg
  \includegraphics[width=4.5in,keepaspectratio]{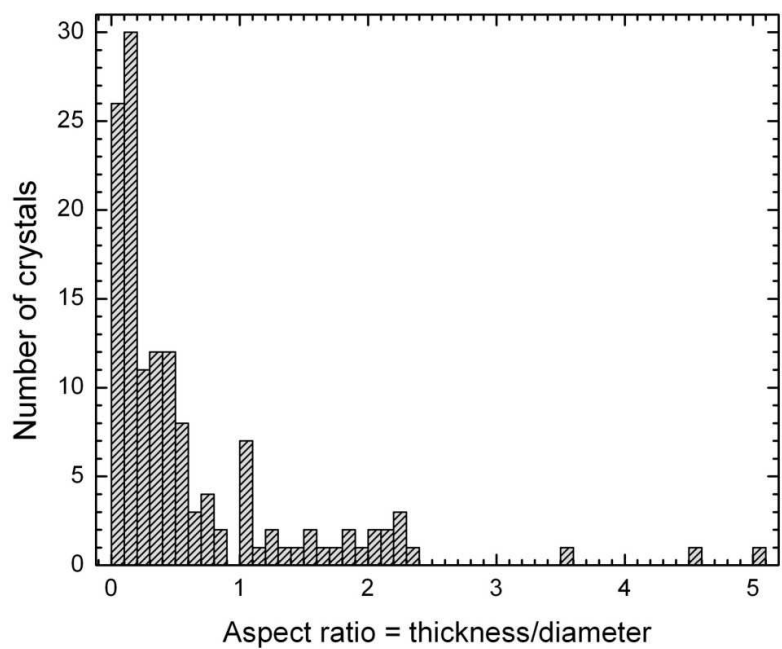}
  \caption{Distribution of aspect ratios of
crystals grown at -8 C, showing a range from plates to columns.}
  \label{d8hist}
\end{figure}

\begin{figure}[tbp] % float placement: (h)ere, page (t)op, page (b)ottom, other (p)age
  \centering
  % file name: C:/Documents and Settings/Kenneth Libbrecht/My Documents/aatempfold/HelenObs/d8x.jpg
  \includegraphics[width=5.5in,keepaspectratio]{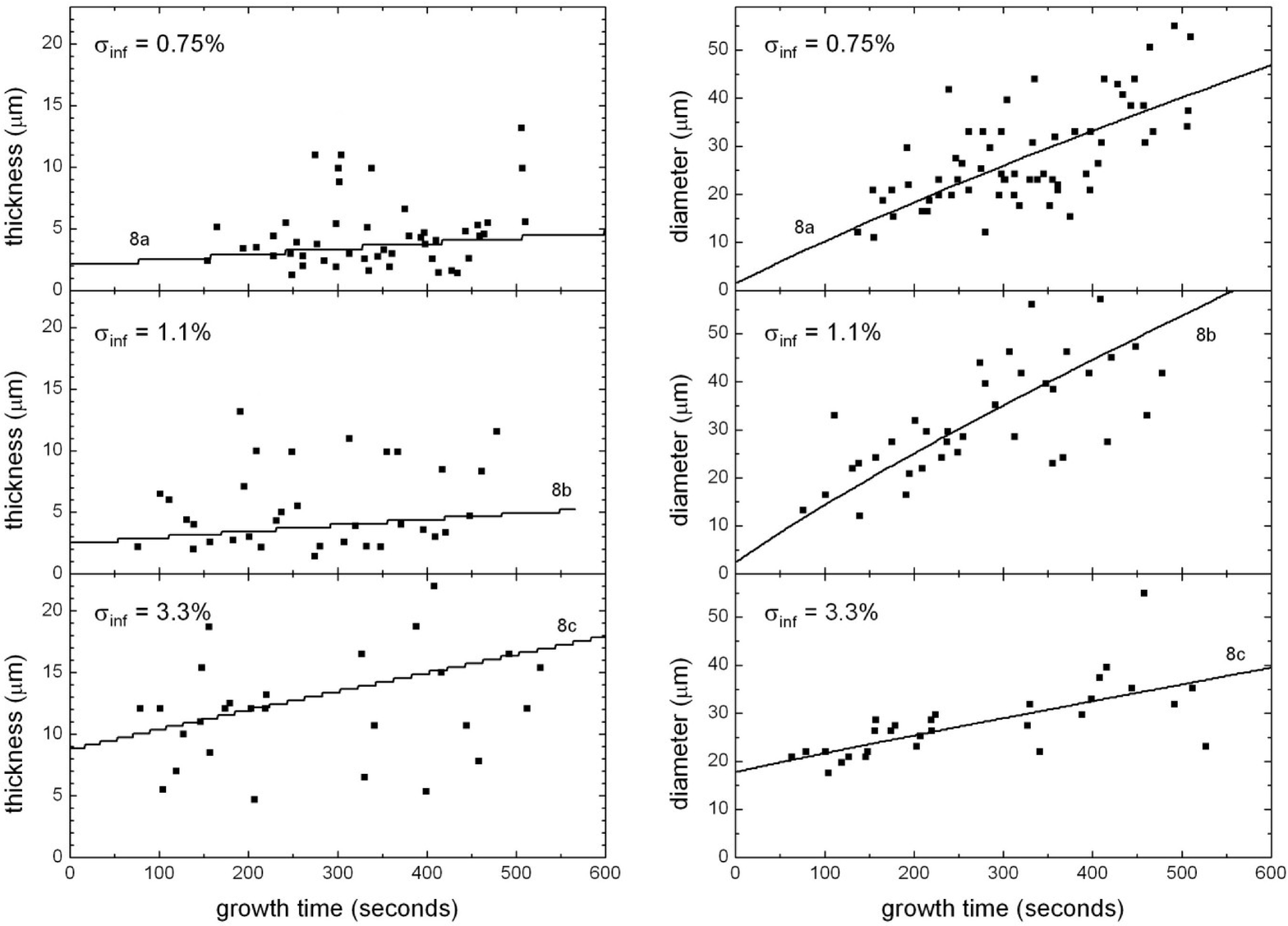}
  \caption{Data from plate-like crystals grown at -8 C, with aspect ratios less
than 0.8, along with model calculations.}
  \label{d8}
\end{figure}

\subsection{T = -8 C}

This temperature is midway between the growth of plate-like and columnar
crystals \cite{libbrechtreview}, so we see much morphological diversity in
the crystals, shown in Figure \ref{d8hist} as a broad distribution of aspect
ratios $A=thickness/diameter.$ In this figure we ignored a small number of
blocky crystals and other poorly formed crystals. Since the majority of the
observed crystals were plate-like, we removed all crystals with aspect
ratios greater than 0.8 (nearly half) to produce the data shown in Figure %
\ref{d8}. In this figure, approximately 25\% of the plates at $\sigma
_{\infty }=0.75\%$ were too thick or too structured to obtain thickness
measurements, while roughly 7\% could not be measured at the higher
supersaturations.

In Figure \ref{d8} we again see a large initial crystal thickness when the
supersaturation is high. The thicker crystals require more mass to increase
the diameter $D,$ with the result that $dD/dt$ is highest at the
intermediate supersaturation of 1.1\%. From the model parameters in the
Appendix, we see that the main change with supersaturation is the initial
crystal diameter $D_{0}.$

\begin{figure}[tbp] % float placement: (h)ere, page (t)op, page (b)ottom, other (p)age
  \centering
  % file name: C:/Documents and Settings/Kenneth Libbrecht/My Documents/aatempfold/HelenObs/d10allx.jpg
  \includegraphics[width=4.5in,keepaspectratio]{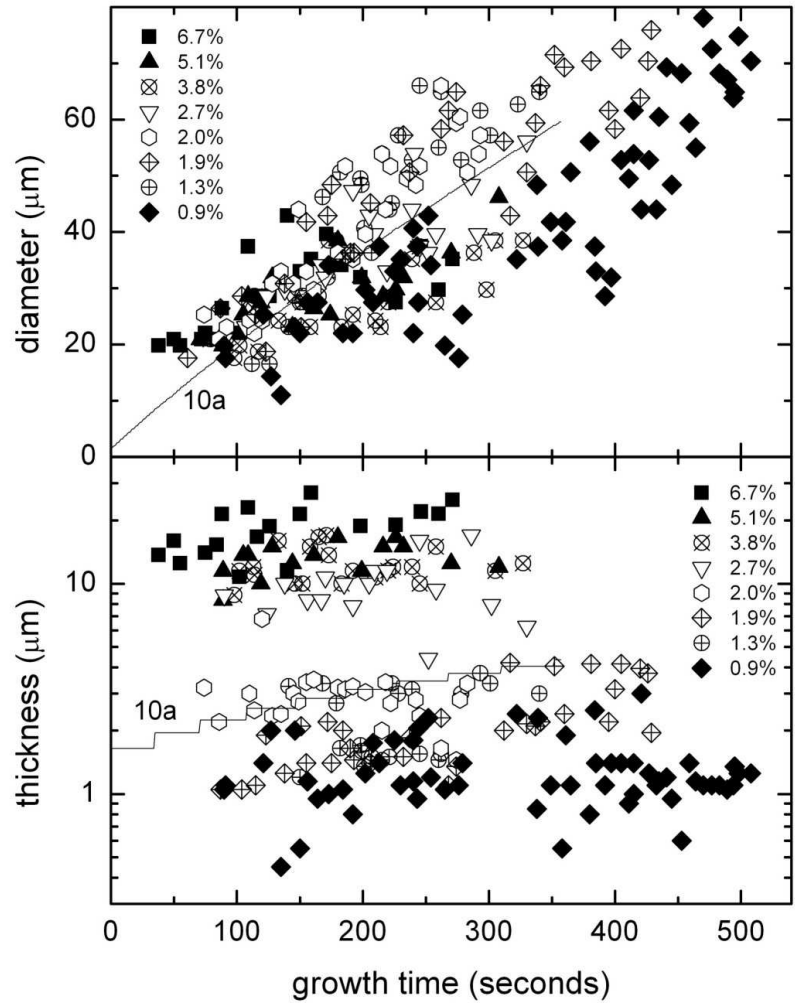}
  \caption{Data taken at -10 C over a
range of supersaturations. Note the clear jump in thickness around $\protect%
\sigma _{\infty }=2.5\%,$ and that in all cases the thicknesses are nearly
constant with time.}
  \label{d10all}
\end{figure}

\subsection{T = -10 C}

At this temperature we took data over a range of supersaturations to examine
the plate thickness as a function of $\sigma _{\infty },$ yielding the data
in Figure \ref{d10all}. Not shown are data taken at 0.7\%, which yielded
essentially all small blocky crystals. At 0.9\% we observed a bimodal
distribution with approximately 20\% blocks (with $0.1<A<1$) and 80\% thin
plates (with $A<0.1$). The plate fraction increased at higher
supersaturations. Over 80\% of the plates had measurable thicknesses, with
no discernible trends in this percentage with $\sigma _{\infty }.$

\begin{figure}[tbp] % float placement: (h)ere, page (t)op, page (b)ottom, other (p)age
  \centering
  % file name: C:/Documents and Settings/Kenneth Libbrecht/My Documents/aatempfold/HelenObs/d10-200x.jpg
  \includegraphics[width=4.5in,keepaspectratio]{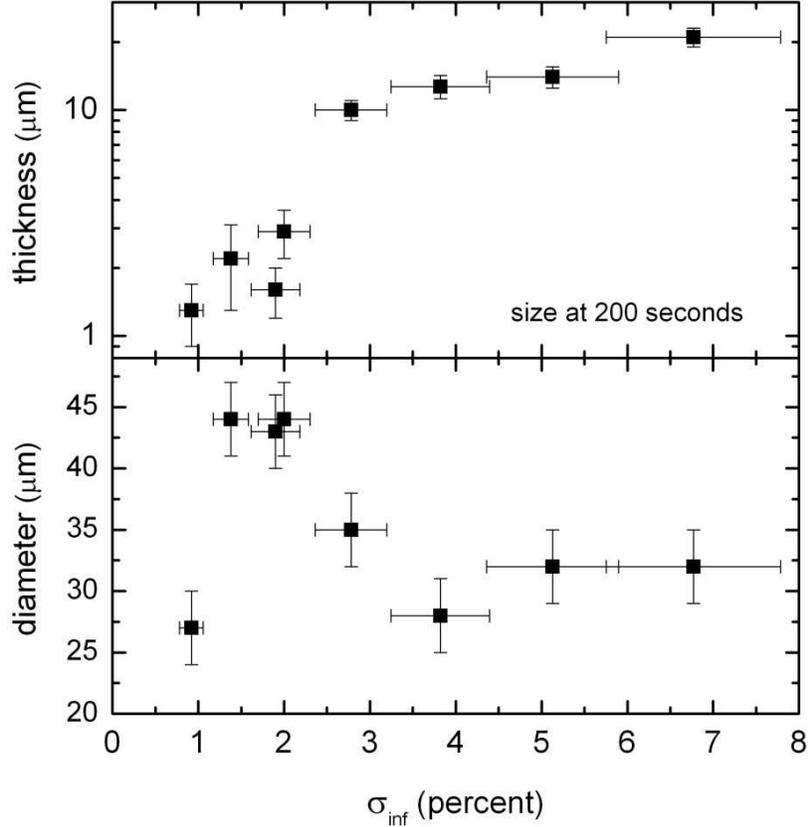}
  \caption{Crystal sizes after a growth time of
200 seconds at -10 C. Note the abrupt jump in thickness at $\protect\sigma %
_{\infty }=2.5\%,$ which is accompanied by a decrease in crystal diameter.}
  \label{d10-200}
\end{figure}

\begin{figure}[tbp] % float placement: (h)ere, page (t)op, page (b)ottom, other (p)age
  \centering
  % file name: C:/Documents and Settings/Kenneth Libbrecht/My Documents/aatempfold/HelenObs/d10modelsx.jpg
  \includegraphics[width=4.5in,keepaspectratio]{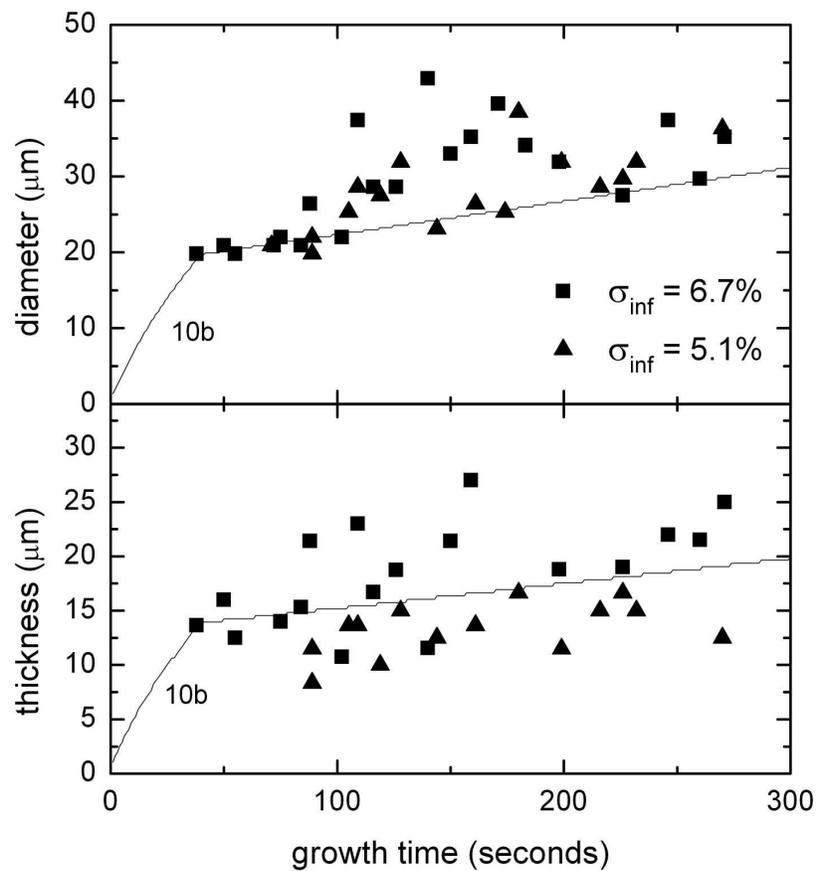}
  \caption{Data at -10 C for the highest supersaturations measured, along
with model calculations described in the text.}
  \label{d10models}
\end{figure}

The -10 C data show the rapid initial growth most clearly, and a distinct
jump in the thickness data can be seen in Figure \ref{d10all}. Note also
that the crystal thickness assumes its nearly constant later value in less
than 50 seconds for the highest supersaturations. Clearly data at shorter
times would be beneficial, and could be acquired by using air currents to
drive crystals quickly onto the substrate. The crystals sizes at 200 seconds
are shown in Figure \ref{d10-200}. Again the thickness shows a rather abrupt
jump around $\sigma _{\infty }=2.5\%.$ We also see a decrease in the
diameter above $\sigma _{\infty }=2.5\%$, which is explained simply by the
jump in thickness, since a thicker crystal requires more mass for a given
diameter.

Figure \ref{d10models} shows the same data at only the highest
supersaturations along with a model calculation in which we assumed one set
of values for $\alpha _{basal}$ and $\alpha _{prism}$ before 40 seconds and
a different set after this time, as listed in the Appendix. These data
suggest that both $\alpha _{basal}$ and $\alpha _{prism}$ decreased with
time as the crystals grew, although other possible explanations have not
been definitely ruled out.

\subsection{T = -12 C}

\begin{figure}[tbp] % float placement: (h)ere, page (t)op, page (b)ottom, other (p)age
  \centering
  % file name: C:/Documents and Settings/Kenneth Libbrecht/My Documents/aatempfold/HelenObs/blocksplatesx.jpg
  \includegraphics[width=4.5in,keepaspectratio]{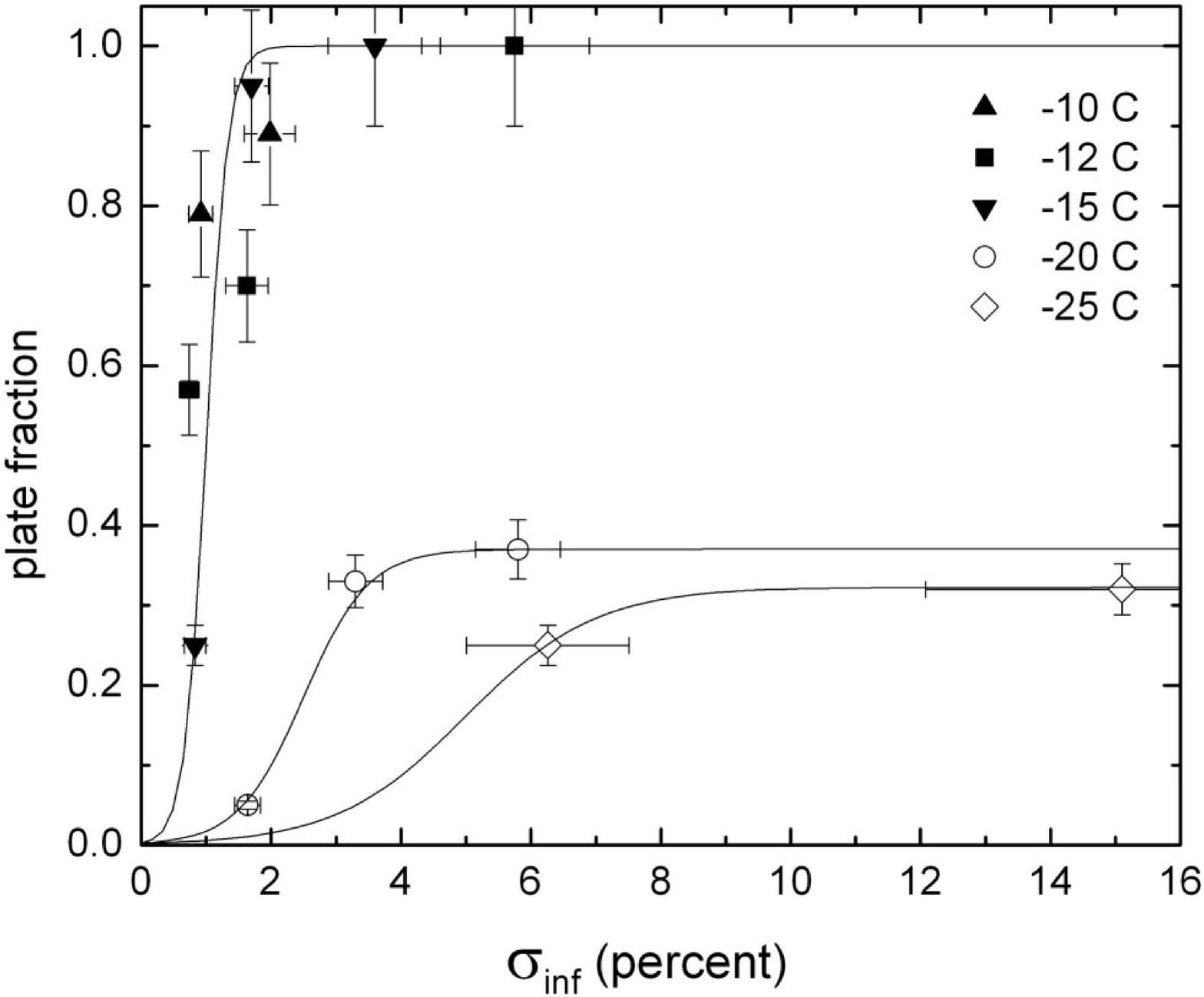}
  \caption{Fraction of crystals that were
plate-like for different temperatures, as a function of supersaturation.
Lines were drawn to guide the eye.}
  \label{blocksplates}
\end{figure}

The crystals at this temperature showed a rather clear transition from
blocky to plate-like morphologies with increasing supersaturation. At $%
\sigma _{\infty }=0.4\%$ we observed no crystals, at 0.7\% about 40\% were
blocky and the rest were plates, at 1.6\% about 30\% were blocks, and at
5.7\% essentially all were plates. Figure \ref{blocksplates} shows the trend
as a function of $\sigma _{\infty }$ for this and other temperatures. The
precise morphologies of the blocky crystals were difficult to determine
since their sizes were quite small and their aspect ratios were close to
unity.

\begin{figure}[tbp] % float placement: (h)ere, page (t)op, page (b)ottom, other (p)age
  \centering
  % file name: C:/Documents and Settings/Kenneth Libbrecht/My Documents/aatempfold/HelenObs/d12x.jpg
  \includegraphics[width=4.5in,keepaspectratio]{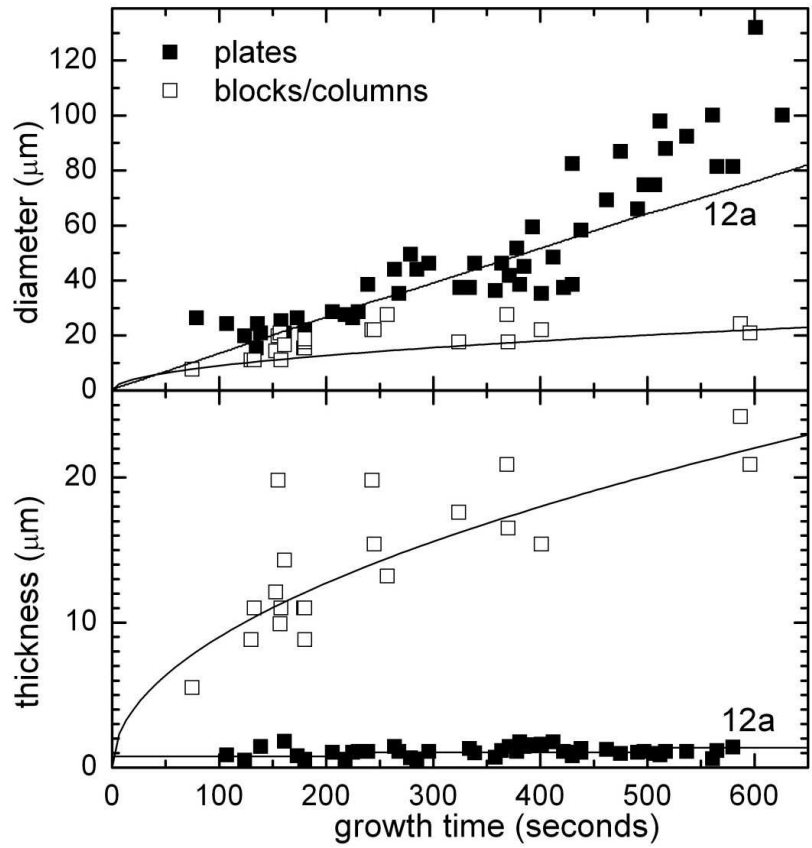}
  \caption{Data taken at -12 C and $%
\protect\sigma _{\infty }=1.6\%,$ along with model calculations.}
  \label{d12}
\end{figure}

\begin{figure}[tbp] % float placement: (h)ere, page (t)op, page (b)ottom, other (p)age
  \centering
  % file name: C:/Documents and Settings/Kenneth Libbrecht/My Documents/aatempfold/HelenObs/d12ax.jpg
  \includegraphics[width=4.5in,keepaspectratio]{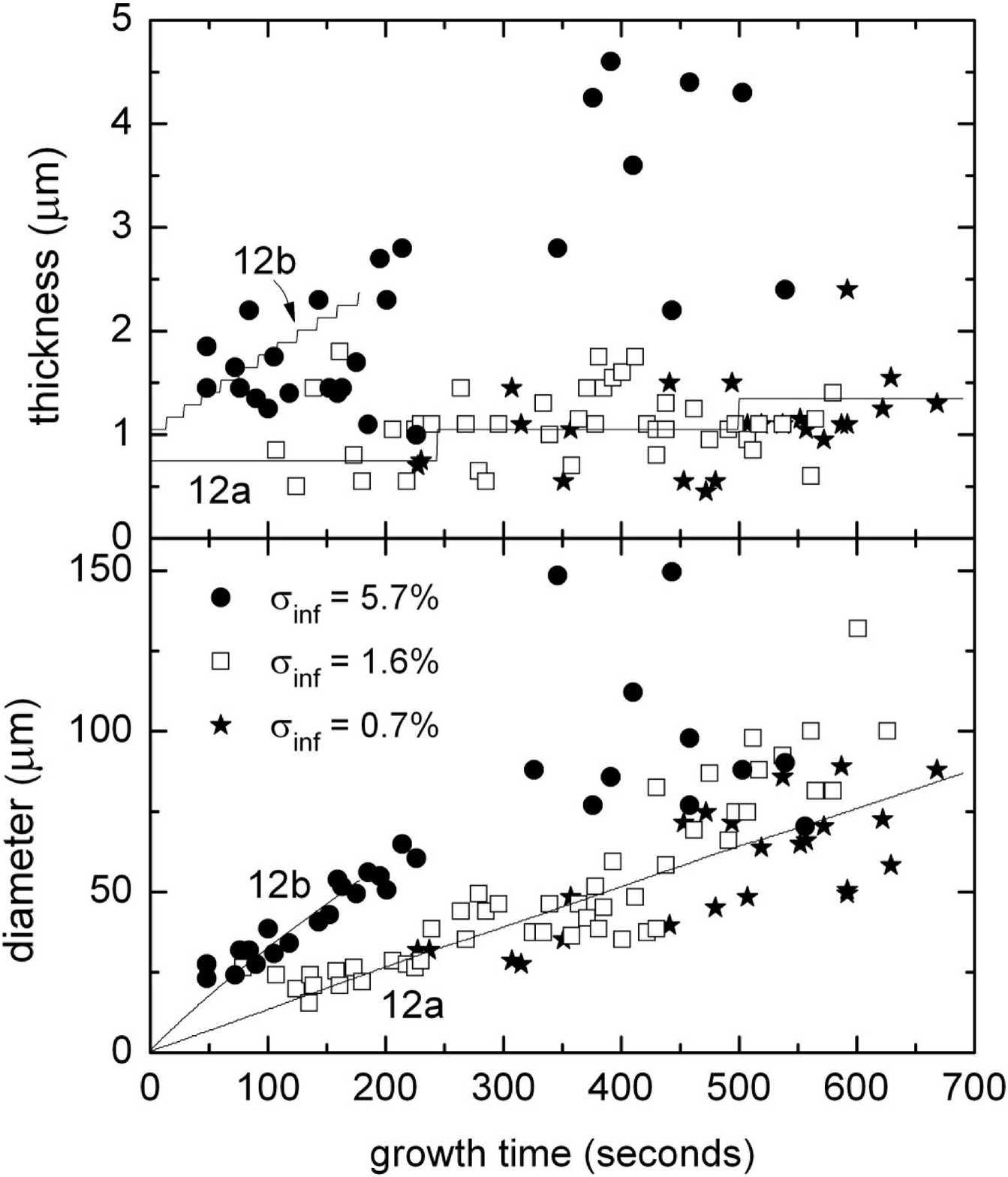}
  \caption{Plate data taken at -12 C, along with model calculations.}
  \label{d12a}
\end{figure}

Figure \ref{d12} shows data at 1.6\% that includes both plates and blocks.
The blocky crystals are reasonably well described with radii that go as $%
R\sim t^{1/2}.$ Since $\alpha \ll \alpha _{diff}$ for these crystals \cite%
{libbrechtreview}, this model implies $\alpha \sim R^{-1},$ again suggesting
that $\alpha $ decreases with growth time even though $\sigma _{surf}$
remains nearly constant. Figure \ref{d12a} shows additional data for plates
grown at -12 C. The plates at this supersaturation exhibited an exceedingly
small $\alpha _{basal}$ (see the Appendix).

\subsection{T = -15 C}

\begin{figure}[tbp] % float placement: (h)ere, page (t)op, page (b)ottom, other (p)age
  \centering
  % file name: C:/Documents and Settings/Kenneth Libbrecht/My Documents/aatempfold/HelenObs/d15alltimex.jpg
  \includegraphics[width=4.5in,keepaspectratio]{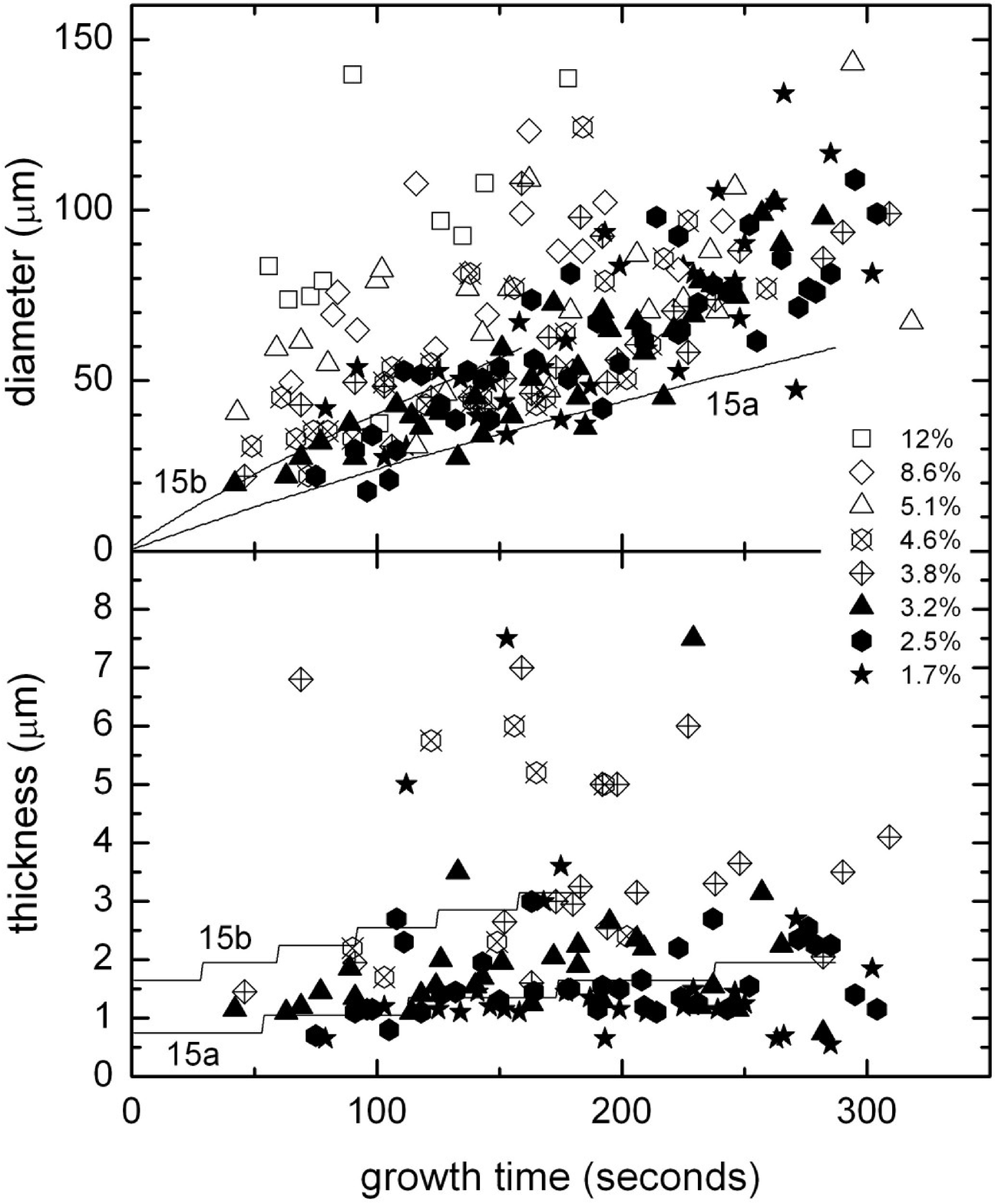}
  \caption{Data taken at -15 C and a
variety of supersaturations, along with two representative models.}
  \label{d15alltime}
\end{figure}

Figure \ref{d15alltime} shows data taken at -15 C over a range of
supersaturations. The complexity of the crystals changed substantially with
supersaturation at this temperature. We observed predominantly simple plates
below $\sigma _{\infty }=3.5\%,$ more structured plates at 5.1\%, simple
branched plates at 8.6\%, and complex branched plates at 12\%. The structure
in the crystals meant that the basal surfaces were often not flat enough to
allow interferometric thickness measurements, as shown in Figure \ref%
{d15comb}. Crystals at this temperature exhibited a marked increase in
thickness with increasing $\sigma _{\infty },$ although with a somewhat
different character than at -10 C. Unfortunately, the structure in the
crystals at -15 C did not allow accurate thickness measurements at high
supersaturations. This temperature also produced the highest $\alpha _{prism}
$ values, consistent with expectations from the morphology diagram.

\begin{figure}[tbp] % float placement: (h)ere, page (t)op, page (b)ottom, other (p)age
  \centering
  % file name: C:/Documents and Settings/Kenneth Libbrecht/My Documents/aatempfold/HelenObs/d15gasesx.jpg
  \includegraphics[width=4.5in,keepaspectratio]{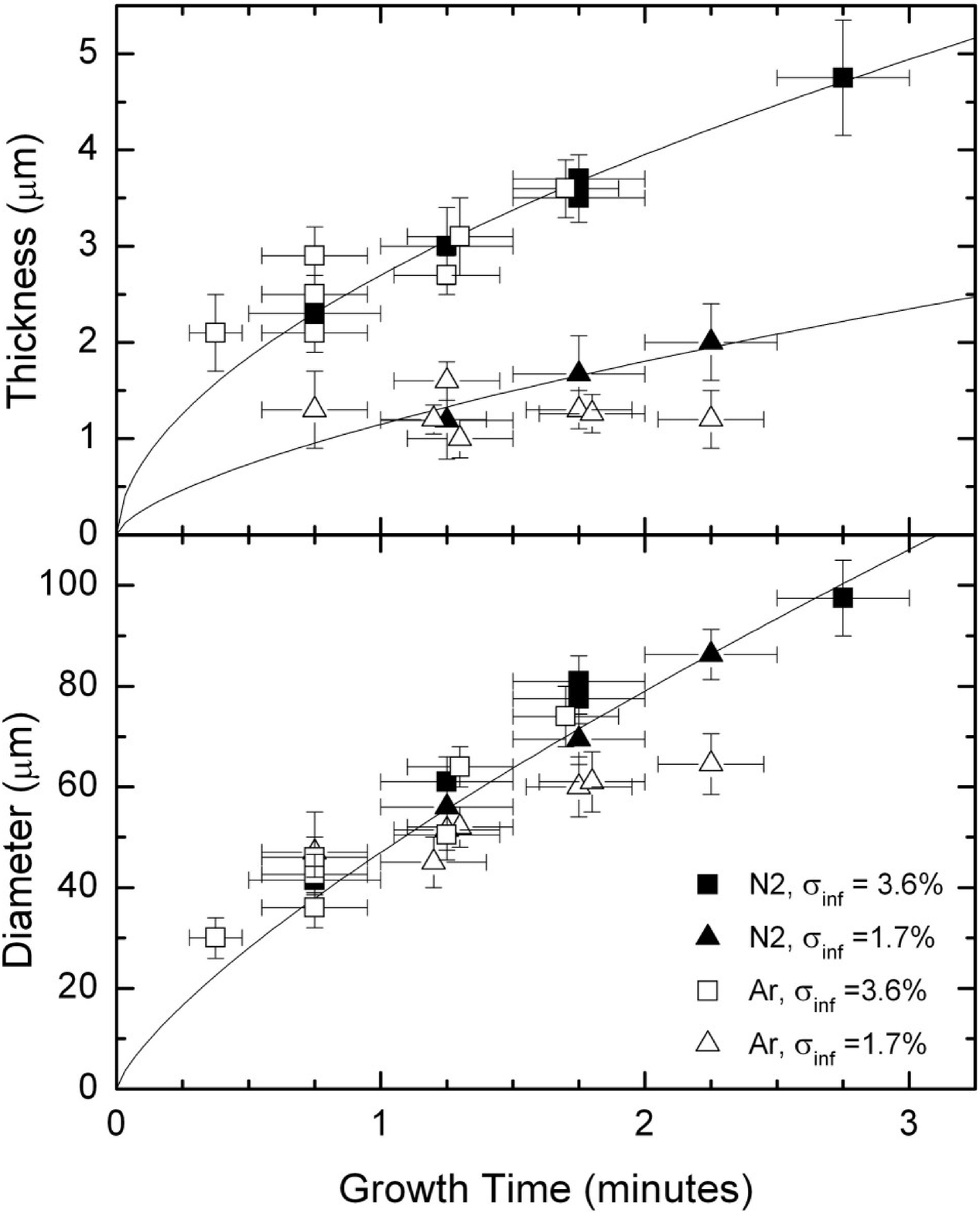}
  \caption{Comparison of crystal growth at
-15 C in nitrogen and argon gas. Lines were drawn to guide the eye. These
data suggest that nitrogen and argon gases are equally inert regarding their
effect on ice crystal growth.}
  \label{d15gases}
\end{figure}

We also examined growth in nitrogen and argon gases at -15 C and a pressure
of one bar, yielding the data in Figure \ref{d15gases}. These data suggest
that nitrogen and argon are equally inert in regards to their effect on ice
crystal growth. A comparison with growth in air (Figure \ref{d15alltime})
further suggests that oxygen is also equally inert.

\begin{figure}[tbp] % float placement: (h)ere, page (t)op, page (b)ottom, other (p)age
  \centering
  % file name: C:/Documents and Settings/Kenneth Libbrecht/My Documents/aatempfold/HelenObs/d15combx.jpg
  \includegraphics[width=4.5in,keepaspectratio]{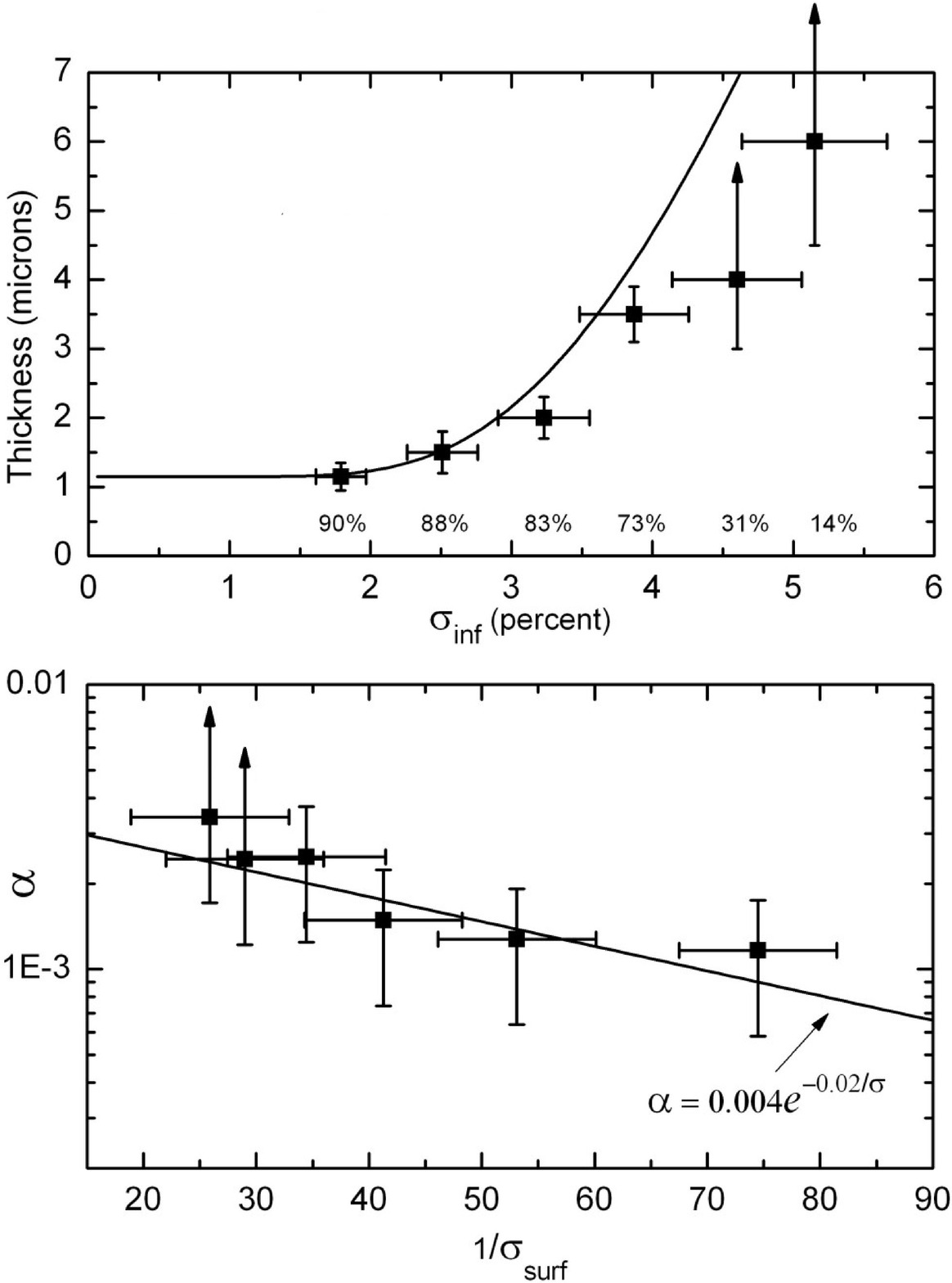}
  \caption{Top panel: Crystal thickness
after 200 seconds of growth at -15 C, as a function of $\protect\sigma %
_{\infty }.$ Numbers below the points give the fraction of crystals with
measurable thicknesses. The line was drawn to guide the eye. Over this same
range in $\protect\sigma _{\infty }$, the diameters at 200 seconds changed
monotonically from approximately 65 to 80 $\protect\mu $m. Lower panel:
Values of the attachment coefficient $\protect\alpha _{basal}$ as a function
of $\protect\sigma _{surf}^{-1},$ as determined from model calculations.}
  \label{d15comb}
\end{figure}

\subsection{T = -20 C}

\begin{figure}[tbp] % float placement: (h)ere, page (t)op, page (b)ottom, other (p)age
  \centering
  % file name: C:/Documents and Settings/Kenneth Libbrecht/My Documents/aatempfold/HelenObs/d20x.jpg
  \includegraphics[width=4.5in,keepaspectratio]{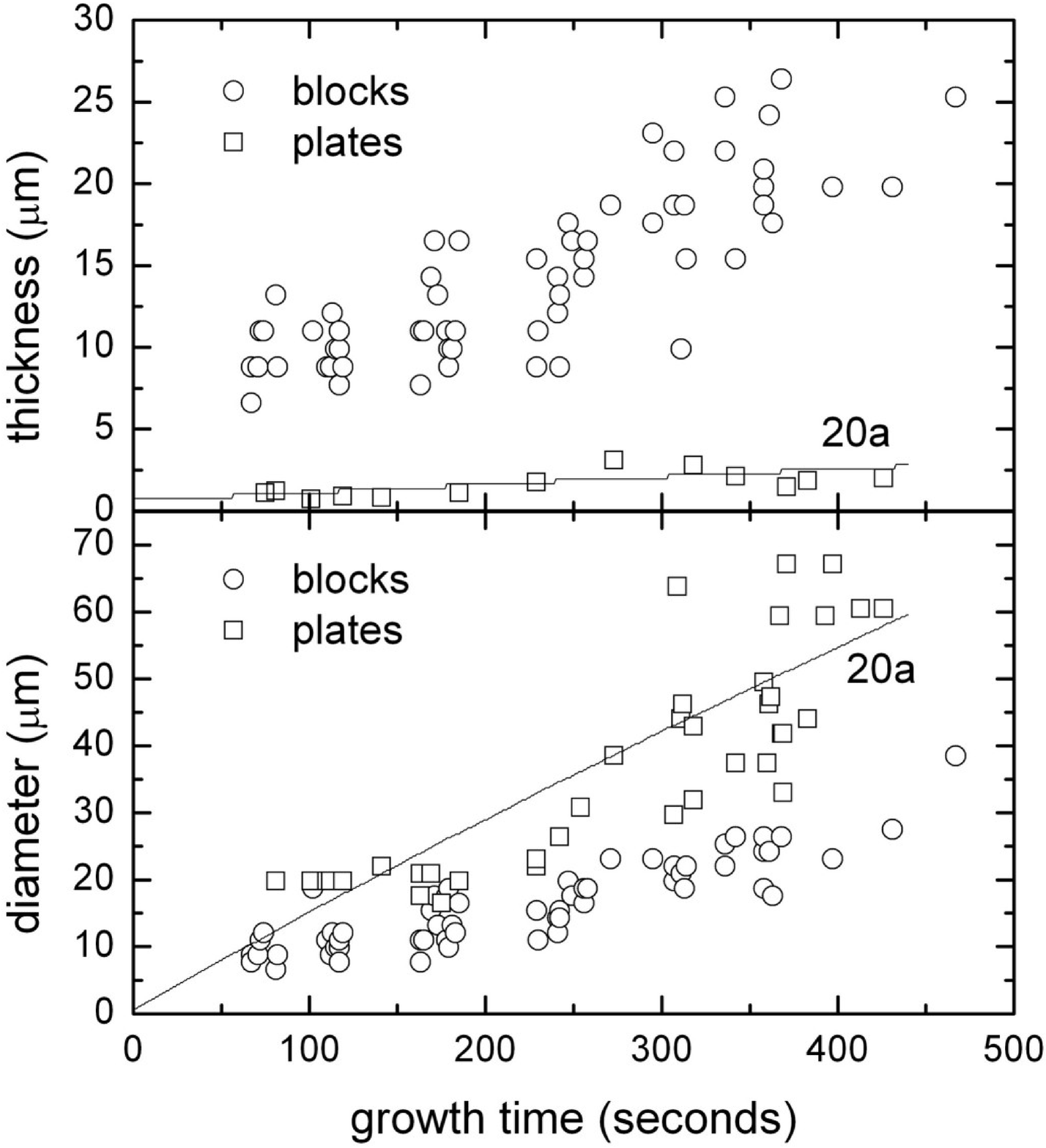}
  \caption{Growth data at -20 C and a
supersaturation of $\protect\sigma _{\infty }=5.8\%,$ showing both
plate-like and blocky crystals.}
  \label{d20}
\end{figure}

We again observed a bimodal distribution of blocky and plate-like crystals
at -20 C. There were few plates at $\sigma _{\infty }=1.6\%$, but more at
higher supersaturations, as shown in Figure \ref{blocksplates}. Growth at
5.8\% is shown in Figure \ref{d20} along with model calculations.

\subsection{T = -25 C}

We did just a few runs at -25 C, and these did not produce much quality
data. At $\sigma _{\infty }=6\%$ the crystals were mainly blocky in form,
with diameters of about 15 $\mu $m at 200 seconds, yielding $\alpha \approx
0.008$ for both the prism and basal faces. At $\sigma _{\infty }=15\%$ a
good fraction of the crystals were plate-like (see Figure \ref{blocksplates}%
), with thicknesses of about 8 $\mu $m and diameters of about 22 $\mu $m at
200 seconds, giving $\alpha _{basal}\approx 0.002$ and $\alpha
_{prism}\approx 0.005.$ At $\sigma _{\infty }=30\%$ the crystals were almost
entirely polycrystalline forms. These data could be improved by nucleating
crystals at a higher temperature and then transferring them to the growth
chamber.

\begin{figure}[tbp] % float placement: (h)ere, page (t)op, page (b)ottom, other (p)age
  \centering
  % file name: C:/Documents and Settings/Kenneth Libbrecht/My Documents/aatempfold/HelenObs/morphx.jpg
  \includegraphics[width=4.5in,keepaspectratio]{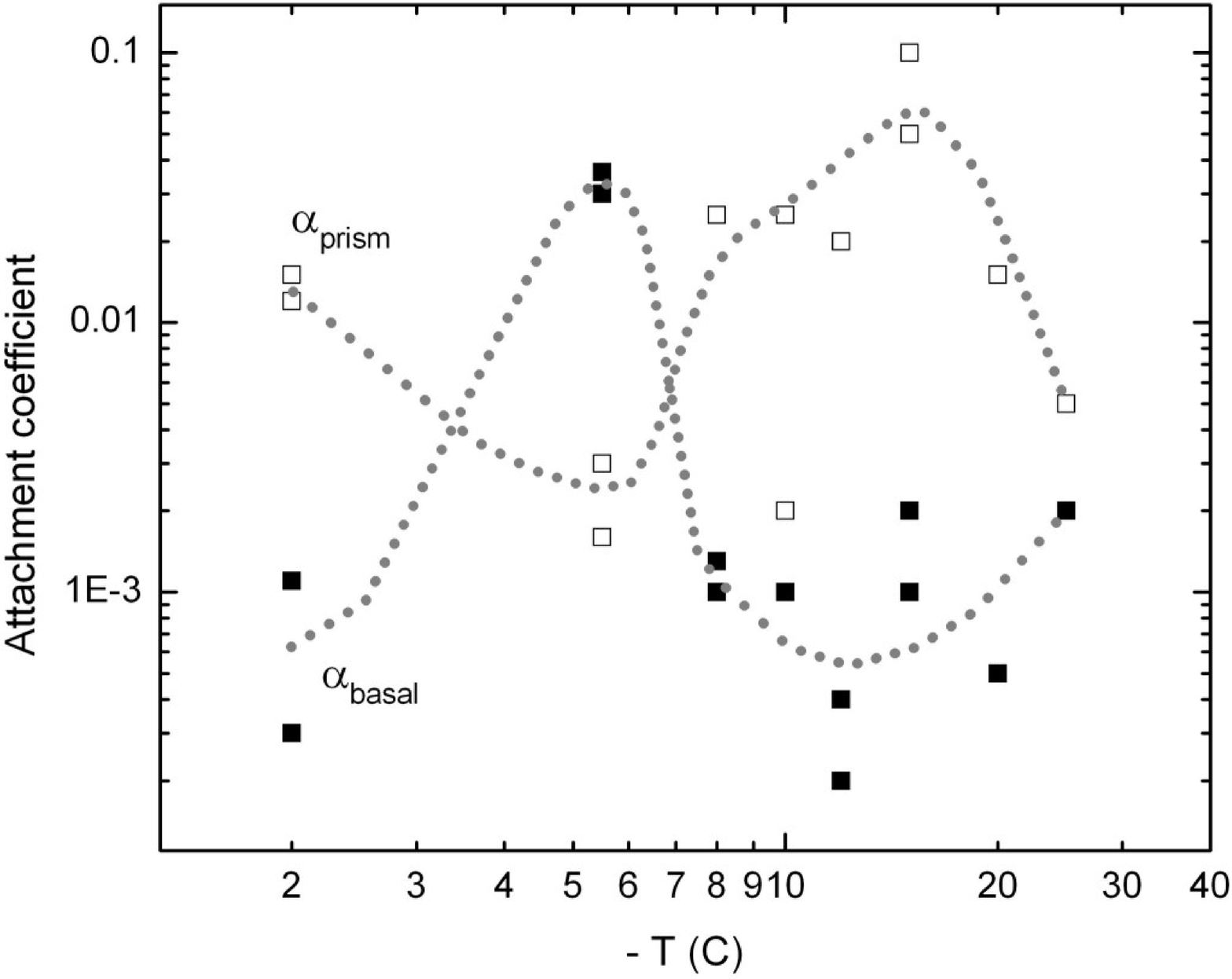}
  \caption{Summary of our measured
attachment coefficients $\protect\alpha _{basal}$ and $\protect\alpha %
_{prism}$ from the Appendix, as a function of temperature. Data at all
supersaturations are shown, and lines were drawn to guide the eye. The
overall trends in these data agree with the well-known snow crystal
morphology diagram.}
  \label{morph}
\end{figure}

\section{Conclusions and Future Work}

As mentioned in the introduction, the data presented here should be taken as
an initial, baseline set of data to be used as a jumping-off point for
future experiments. At present there is little quantitative theory with
which we can compare the $\alpha $ values we determined. The crystal
morphologies we observed as a function of temperature and supersaturation
were in general agreement with the known morphology diagram, as were the
extracted attachment coefficients shown in Figure \ref{morph}. By modeling
the growth of isolated crystals, our observations are a step toward
replacing the qualitative morphology diagram with more quantitative
measurements of $\alpha _{basal}$ and $\alpha _{prism}$ as a function of
temperature and supersaturation.

The present measurements were made mainly in ordinary air, which contains
impurities at some unknown level. We now believe that gaseous impurities
play an essential role in snow crystal growth \cite{impurities}, which means
much greater attention must be paid to the composition of the background
gas. Additional experiments with gaseous impurity levels both higher and
lower than we had in the current experiments would be especially
interesting. The data presented here are useful for determining optimal
strategies for investigating ice crystal growth in different chemical
environments.

We observed a number of features in our data that should be investigated
further. The fast initial growth observed at -10 C (as well as at other
temperatures) may be a useful indicator of impurity effects, and this is
certainly worth examining further, especially as a function of
supersaturation. Statistics on the numbers of triangular plates produced, as
well as other unusual plate-like morphologies, could also be pursued. A more
careful examination of growth morphologies between -6 C and -10 C may also
yield better insights into the transition from plate-like to columnar growth.

\section{Acknowledgements}

We acknowledge support for HCM and BF by the Caltech Summer Undergraduate
Research Fellowship (SURF) program, the CamSURF program, and the Robert L.
Blinkenberg SURF Endowment.

\section{Appendix - Model Parameters}

We modeled our data using the 2D cellular automata method described in \cite%
{libbrechtmodel}. We used constant input growth parameters $\alpha _{basal}$
and $\alpha _{prism},$ along with an input initial crystal diameter $D_{0}.$
These parameters were adjusted to to produce model crystals that fit our
measurements. In addition to the crystal thickness and diameter as a
function of time shown in the graphs, the model also yielded $\sigma _{surf}$
at all points on the crystal surface as a function of time. From these data
we extracted roughly average values for $\sigma _{surf}$ on the basal and
prism facets, which are shown in the table. We believe that overall
uncertainties in $\sigma _{\infty }$ and the derived quantities are roughly
20\% throughout, mainly from a variety of systematic errors in the
measurements.

\vskip10pt

\begin{tabular}{cccccccc}
Model\# & $T$ (C) & $\sigma _{\infty }$ (\%) & $\alpha _{basal}$ & $\alpha
_{prism}$ & $D_{0}$ ($\mu $m) & $\sigma _{surf,basal}$ (\%) & $\sigma
_{surf,prism}$ (\%) \\ 
2a & -2 & 1.1 & 0.0003 & 0.015 & 1 & 1.0 & 0.9 \\ 
2b & -2 & 2.4 & 0.0011 & 0.012 & 3 & 2.0 & 1.8 \\ 
5a & -5.5 & 0.5 & 0.03 & 0.0016 & 2 & 0.4 & 0.45 \\ 
5b & -5.5 & 1.1 & 0.036 & 0.003 & 3 & 0.8 & 0.95 \\ 
5c & -5.5 & 4.4 & 0.03 & 0.003 & 7 & 3.0 & 3.5 \\ 
8a & -8 & 0.75 & 0.0013 & 0.025 & 2 & 0.6 & 0.5 \\ 
8b & -8 & 1.1 & 0.001 & 0.025 & 3 & 0.9 & 0.75 \\ 
8c & -8 & 3.3 & 0.001 & 0.025 & 18,9 & 2.6 & 2.1 \\ 
10a & -10 & 2 & .001 & .025 & 2 & 1.6 & 1.5 \\ 
10b & -10 & 6 & 0.02,0.001 & 0.03,0.002 & 1 & 4.0 & 4.0 \\ 
12a & -12 & 1.6 & 0.0002 & 0.02 & 1 & 1.4 & 1.3 \\ 
12b & -12 & 5.7 & 0.0004 & 0.02 & 1 & 5.0 & 4.0 \\ 
15a & -15 & 2 & 0.001 & 0.1 & 1 & 1.5 & 1.2 \\ 
15b & -15 & 4 & 0.002 & 0.05 & 2 & 3.0 & 2.2 \\ 
20a & -20 & 5.8 & 0.0005 & 0.015 & 1 & 5.2 & 5.0%
\end{tabular}

\end{document}